\documentclass[]{heart2014_4Pre}

\title{A Reconfigurable Vector Instruction Processor for Accelerating a Convection Parametrization Model on FPGAs}

 \numberofauthors{3}
 \author{ 
 	\alignauthor Syed Waqar Nabi\titlenote{Also affiliated with: \\ Namal College, Mianwali, 42250, Pakistan.}\\
 	\affaddr {School of Computing Science,}\\
 	\affaddr {University of Glasgow,} \\
 	\affaddr {Glasgow G12 8QQ, UK.}\\ 
 	\email {syed.nabi@glasgow.ac.uk}
 	\alignauthor Saji N. Hameed \\ 
 	\affaddr {University of Aizu,}\\ 
 	\affaddr {Tsuruga, Ikki-machi,} \\
 	\affaddr {Aizuwakamatsu-shi, Japan.}\\
 	\email {saji@u-aizu.ac.jp}\\
 	\alignauthor Wim Vanderbauwhede\\ 
 	\affaddr {School of Computing Science,}\\
 	\affaddr {University of Glasgow,} \\
 	\affaddr {Glasgow G12 8QQ, UK.}\\ 
 	\email {wim@dcs.gla.ac.uk}\\
 }

\begin{document}

\maketitle

\begin{abstract}
High Performance Computing (HPC) platforms allow scientists to model computationally intensive algorithms. HPC clusters increasingly use General-Purpose Graphics Processing Units (GPGPUs) as accelerators; FPGAs provide an attractive alternative to GPGPUs for use as co-processors, but they are still far from being mainstream due to a number of challenges faced when using FPGA-based platforms. Our research aims to make FPGA-based high performance computing more accessible to the scientific community. In this work we present the results of investigating the acceleration of a particular atmospheric model, Flexpart, on FPGAs. We focus on accelerating the most computationally intensive kernel from this model. 
The key contribution of our work is the architectural exploration we undertook to arrive at a solution that best exploits the parallelism available in the legacy code, and is also convenient to program, so that eventually the compilation of high-level legacy code to our architecture can be fully automated. 
We present the three different types of architecture, comparing their resource utilization and performance, and propose that an architecture where there are a number of computational cores, each built along the lines of a vector instruction processor, works best in this particular scenario, and is a promising candidate for a generic FPGA-based platform for scientific computation. We also present the results of experiments done with various configuration parameters of the proposed architecture, to show its utility in adapting to a range of scientific applications.

\end{abstract}

\section{Introduction}

Atmospheric and climate scientists are one of the most important users of high-performance computing (HPC) platforms. Models predicting global weather and climate, or more localized phenomenon like after-effects of the Fukushima nuclear disaster or the Icelandic volcano eruption, require the execution of computationally intensive models. 

Today's HPC systems use massive clusters of multi-core CPUs. Increasingly, GPGPUs are being utilized as co-processors to off-load computationally intensive kernels, as they can offer much greater thread-level parallelism at a lower power budget. 

Field-Programmable Gate Arrays (FPGAs) are another viable option for accelerating scientific models by working as co-processors that off load computationally intensive kernels in a manner similar to the way it is done with GPUs. FPGAs moreover can achieve the same performance at much lower power budgets than either CPU only or CPU-GPU systems, due to their ability to match hardware to the algorithm. However, the scientific community in general has been reluctant to use FPGA-based HPC systems due to the challenges associated with programming and exploiting the parallelism available in an FPGA, and also due to poor portability and standardisation \cite{vanderbauwhede2013high}.  

Our work is motivated by the need to make FPGA based HPC systems more accessible to the scientific community. We propose an FPGA-based architecture in the context of a particular atmospheric particle dispersion model, Flexpart, with the view to map the various granularities of parallelism available in a typical scientific model to the parallelism available in the FPGA hardware. In addition, we aim to make this architecture  straight-forward to program using a high-level language and eventually achieve fully automated compilation of legacy code that was originally written for single-core CPUs to our FPGA system.


%

\section{FPGAs for Accelerating Scientific Computation}

FPGAs are inherently parallel architectures that allow compile-time reconfiguration of the hardware through Hardware Description Languages (HDLs) like Verilog. While the overhead of this flexibility results in them being less power and area efficient than Application-Specific ICs (ASICs), they can be much more power-efficient than instruction processors like CPUs and GPUs. This is because FPGAs can be tailored to the algorithm that is being executed, provide parallelism at various granularities, and can give comparable overall throughput at much lower frequencies. 
So larger, faster systems can be built with lower power-budgets. 


There are a number of challenges associated with using FPGAs for HPC. FPGAs, like any other co-processor, are limited by the IO bandwidth due to the loose coupling with the host CPU, typically over an interface like PCI Express. This constraint places and upper limit on the achievable speed up from any acceleration co-processor. FPGAs are hence suitable for accelerating small pieces of code -- the  kernels -- that take the most amount of time to execute. 


Programming FPGAs is a very different paradigm from programming instruction processors, and it is generally considered outside the domain of software programmers. There has been considerable progress on this front in recent years, for example the HLS tools of Xilinx\cite{105.100}, and Altera'a venture into using OpenCL for programming FPGAs \cite{105.101}. However, it can safely be said that this field is far from mature, and lagging considerably behind the kind of tools and support available to programmers of CPUs and GPUs. 

It is not trivial to convert legacy code for implementation on FPGAs even with the introduction of high-level programming paradigms. Porting an existing CPU-based code to a HPC platform based on FPGAs may in fact even be more difficult than writing completely new code for that FPGA-based system. Poor portability and a lack of standardization adds to the challenge and has lead to a reluctance in the scientific community to use FPGAs for their HPC needs.


In view of these challenges, our work looks at atmospheric models -- specifically the Flexpart particle dispersion model -- and we investigate a number of architectural options that could be used to 1) effectively exploit parallelism at various granularity levels available in an algorithm and 2)to do it in a way that makes it easier to program the FPGA, and eventually can lead to a fully automated compiler.

\section{The Flexpart Lagrangian Particle Dispersion Model}

Flexpart is a meso-scale atmospheric particle dispersion model \cite{104.102}. It can model the transport and diffusion of \textit{particles} -- where particles can be tracers in the air or packets of air itself. The term \textit{meso-scale} means it covers phenomena in an area of somewhere between 5 kilometres to several hundred kilometres. Applications of models at this scale can include phenomena like transport of radio-nuclide, pollution transport, greenhouse gas cycles etc.  
 

\subsection{Lagrangian and Eulerian Models}

Flexpart is a \textit{Lagrangian} particle dispersion model. This is one of the two approaches typically taken in fluid dynamics for modelling the flow of fluids, the other being \textit{Eulerian} modelling. 

A Lagrangian model models the trajectory of a large number of \textit{particles} for every time step. In Flexpart for example, each particle is  moved over time through a 3-dimensional space, and the final result is produced by counting the particles in each grid-box. 

An Eulerian model on the other hand models the mass flux of fluid  and its change over time at grid boundaries. 

A Lagrangian model is more accurate as it accurately models the trajectory of each individual particle, which is not effected by the grid-size. The grid is only applied at the output. In a Eulerian model, a particle released from a point source instantaneously looses its position in the grid-box (this is called Numerical Diffusion). A Lagrangian model is good at representing narrow plumes, in which case its computation requirements can be quite modest as well. However, in general a Lagrangian model would be more computationally intensive than a Eulerian model. 

\subsection{The Working of Flexpart}

\begin{figure}[h]
\centering
\includegraphics[width=0.9\linewidth]{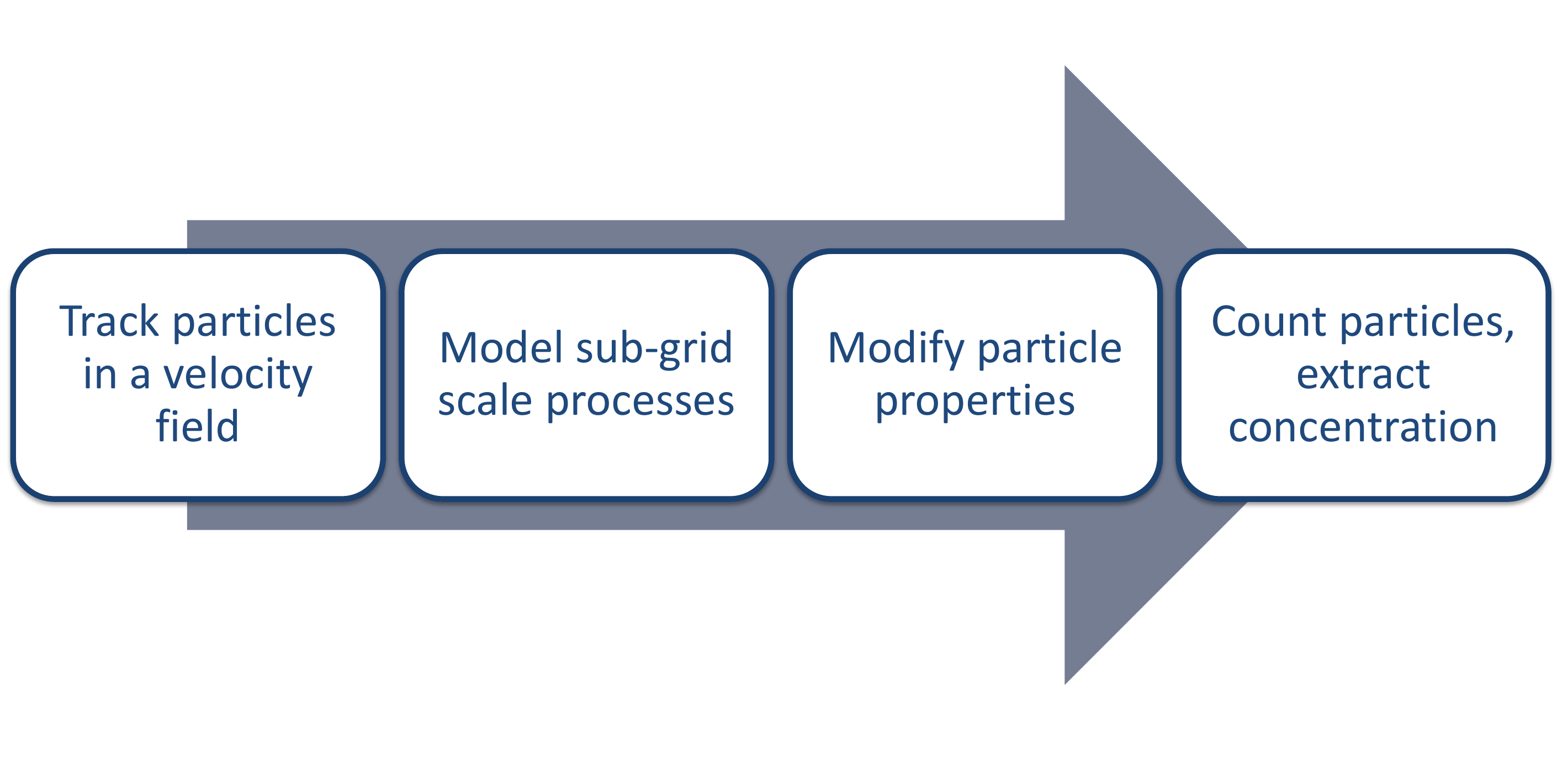}
\caption[]{The Working of the Flexpart Particle Dispersion Model}
\label{fig:flexpart_working}
\end{figure}

Flexpart uses Lagrangian dynamics to model particle dispersion at the meso-scale. Figure \ref{fig:flexpart_working} shows the main steps in the Flexpart model.

Flexpart uses meteorological fields taken from the ECMWF numerical weather prediction model data \cite{104.103} on a latitude / longitude 60-km resolution grid. 
The number of levels in the vertical can be up to 60. This data is interpolated and used by the model to track the trajectory of individual particles over time in a given velocity field.   

There are some phenomena, e.g. convective transport, which are not captured by this grid-scale modelling, and they are called \textit{sub-grid scale} phenomena. Flexpart uses parametrization schemes for modelling the effect of these phenomena on the dispersion of particles. 
As we will see in the next section, the parametrization of convective transport places a significant computational burden on the processor.

If required, 
the individual particle properties can me modified, for example, loss of mass through radioactive decay. 

These three steps are repeatedly executed, and the output is produced at the required time intervals or at the end of simulation by counting particles and extracting the concentration for each grid-box.

\subsection{The Flexpart Source Code}

The Flexpart model has been written in Fortran 77. It is relevant to note that the author did not make any attempts to parallelise the code as the model is strictly sequential \cite{104.102}. 

These meteorological inputs to Flexpart from the ECMWF weather model are available at 3-hour time intervals, whereas the flexpart integration step is typically much smaller, for example 60 seconds in case of our experiments. Their interpolation forms one key element of computation in Flexpart.

The other key method is the modelling of \textit{dispersion}, which moves each particle through a certain distance for that integration time step. 

As discussed earlier, sub-grid scale phenomena like convective transport of particles cannot be modelled by trajectory calculations, and have to be parametrized. Our current work focuses on acceleration of the parametrization algorithm used for modelling the effects of convective turbulence.

\subsection{The Convect Method}

Flexpart uses the scheme proposed by Emmanuel and Zivkovic-Rothman\cite{104.104} for parametrizing the effect of convection on the position of individual particles. 
The model uses grid-scale data available from the ECMWF model and produces a displacement probability matrix, giving the required mass flux information based on which particles can be redistributed.

The Convect kernel is called once per grid-column for each simulation time step. While subsequent calls to Convect for the same grid-column must be executed sequentially due to a data dependency, the calls to Convect for different grid-columns in the same time step are independent of each other. The Convect kernel requires a number of computational steps that must be executed in a linear fashion, with data-dependencies from one step to the other. These observations form the basis of our proposed solution to parallelize the execution of this method on FPGAs.


\subsection{Analysis of Flexpart and Convect for Acceleration}

Flexpart has already been ported for GPU acceleration, and a 10$\times$ performance improvement on a NVIDIA’s Tesla C1060 has been reported \cite{104.018}. However, this comparison excludes the Convect kernel. The authors of Flexpart indicate that the Convect kernel consumes a significant proportion of the overall computation time \cite{104.102}, hence our work focused on this method for acceleration on FPGAs. The protracted nature of the Convect kernel makes it an inconvenient algorithm to tackle, but it also makes it a good representative scientific model that can be used for investigating acceleration of scientific models in general, which is our long-term aim.

Initial simulation seemed to indicate that the execution of Convect has a proportionately negligible contribution to the overall processing time. For example, in a 12-hour simulated-time run, the Convect kernel takes a mere 0.06\% of the 657 seconds of simulation time (See Table \ref{tab:convect_perspective}). 

\begin{table}
\begin{tabular}{|p{25ex}|p{6ex}|p{6ex}|p{6ex}|p{6ex}|}
\hline \textbf{Simulated time (hrs)} 						& 03 hrs & 12 hrs & 120 hrs & 360 hrs  \\ 
\hline \textbf{Simulation time (sec)}						&  192 & 657  & 5,693 & 27,150  \\ 
\hline\textbf{ Synch. steps }								&  180 & 720  & 7200 & 21,600 \\ 
\hline \textbf{Calls to Convect per step (thousands)}       &  0.5 & 6.2 & 3,558 & 66,299 \\ 
\hline \textbf{Average calls per time step }					&  2.7 & 8.6  & 464 & 3,069 \\ 
\hline \textbf{Percentage compute time taken by Convect} 	&  0.02 & 0.06 & 7.55 & 35 \\ 
\hline 
\end{tabular}
\caption{Putting Convect in Perspective}
\label{tab:convect_perspective}
\end{table}

Longer simulation runs however indicated that calls are made to the Convect kernel increasingly as the simulated time progresses, as can be seen from the fourth row in Table \ref{tab:convect_perspective}. Convect -- being a parametrisation of the actual physical phenomena -- is not called on a per-particle basis, but on a per-grid-column basis. So it only needs to be invoked for those vertical columns of air where particles are present. Initially, the particles are typically localised into few grid-columns. As time progresses, due to the dispersion, particles spread out and Convect needs to be called more and more often for each time-step. By extrapolating our results, we see that if Convect were to be called for every vertical column of air in every time step right from the beginning of the simulation, it would take up to 80\% of the entire simulation time. 
Consequently, we focused on porting the Convect kernel for FPGA acceleration.

\section{Architectural Exploration}

\subsection{Design Considerations}

The application domain of scientific computing in general requires considerable arithmetic operations on large arrays of numbers. So there are critical loops performing arithmetic operations on large data, and the flexibility of FPGAs allows us to deal with these loops in a number of different ways, as our following experiments will show.






Identification of parallelism of the appropriate granularity is a very important consideration for accelerating any code on a HPC system. FPGA can be configured to offer various granularities of parallelism, ranging from instruction-level, data-level to thread-level. Hence we can map the architecture to the parallelism exposed by the algorithm rather than trying to do it the other way round. 



In a single call to Convect, the sequence of processes have data-dependencies from one process to the other. Repeated calls to convect in different time-steps for the same grid-column also have a data-dependency. However, the calls to convect in a particular time step for different grid-columns can be executed completely independent of and parallel to each other.

We have explored three architectural alternatives for implementing a Convect kernel on FPGAs. The common feature in all of these architectures is that they are meant to be replicated on an FPGA to make use of the fact that calls to Convect for different grid-columns are independent of each other and can run in parallel. The three alternatives differ in the amount of parallelism that we expose inside a single call to Convect. 

The first two architectures are what may be called ``naive'' approaches; they provide a baseline at two extremes, and lead towards the proposed third option. 

In all three architectures, we conservatively use 64-bit fixed-point numbers, with 32 bits for the fractional part. Since the data available from ECMWF models is in floating-point, we instantiate a floating-point to fixed-point converter in the FPGA for dynamic conversions in all three cases \cite{109.031}. The slice usage and calculation latency estimates are shown for Virtex-5 XC5VLX devices. 

\subsection{Option 1: Fully Tiled Implementation}

The first architecture is a baseline implementation of a small sub-set of the Convect kernel, which computes an intermediate result used subsequently. We unroll the loop, and dedicate a hardware unit for each arithmetic operation. A single iteration of the loop involves 6 multiplications, two divisions, two additions, and one inversion. 

\begin{figure}[h]
\centering
\includegraphics[width=1.0\linewidth]{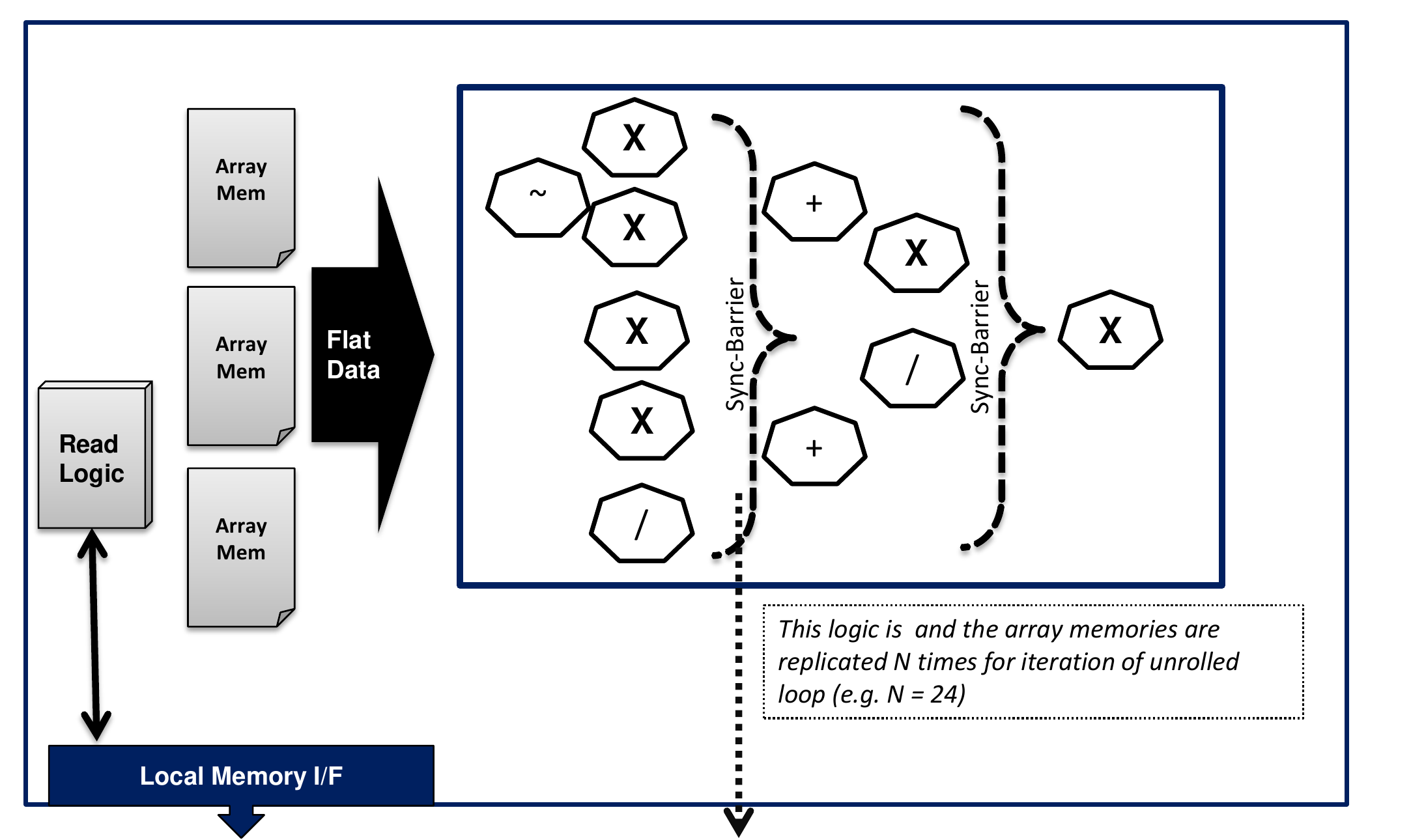}
\caption{First architectural option: fully unrolled and tiled implementation}
\label{fig:option1}
\end{figure}

Like most loops in the Convect kernel, this one loops over each vertical level of air in that grid-column, which numbered 24 in our simulation runs. 
We replicated the compute-unit 24 times. See Figure \ref{fig:option1} for a block diagram of this stage. Note that it does not require any controller to sequence the operations as the algorithm has been \textit{programmed} into the circuit. A synchronisation barrier is required as different arithmetic units can have different computational latencies. 

This approach would require a unique and custom stage for each major loop in the Convect kernel, and as our results will show, it is not a feasible option. However, the architecture serves as an interesting baseline to compare our other approaches.  

\subsection{Option 2: Fully Sequential Implementation}
\label{sec:option2}

Our second approach explores the other extreme of reusing minimum arithmetic units sequentially to compute the loop. This requires a controller and a way to communicate the sequence of operations to it, which would be in the form of instructions stored in a memory. Thus our design approaches that of a typical --  very basic -- instruction processor. 

\begin{figure}[h]
\centering
\includegraphics[width=1.0\linewidth]{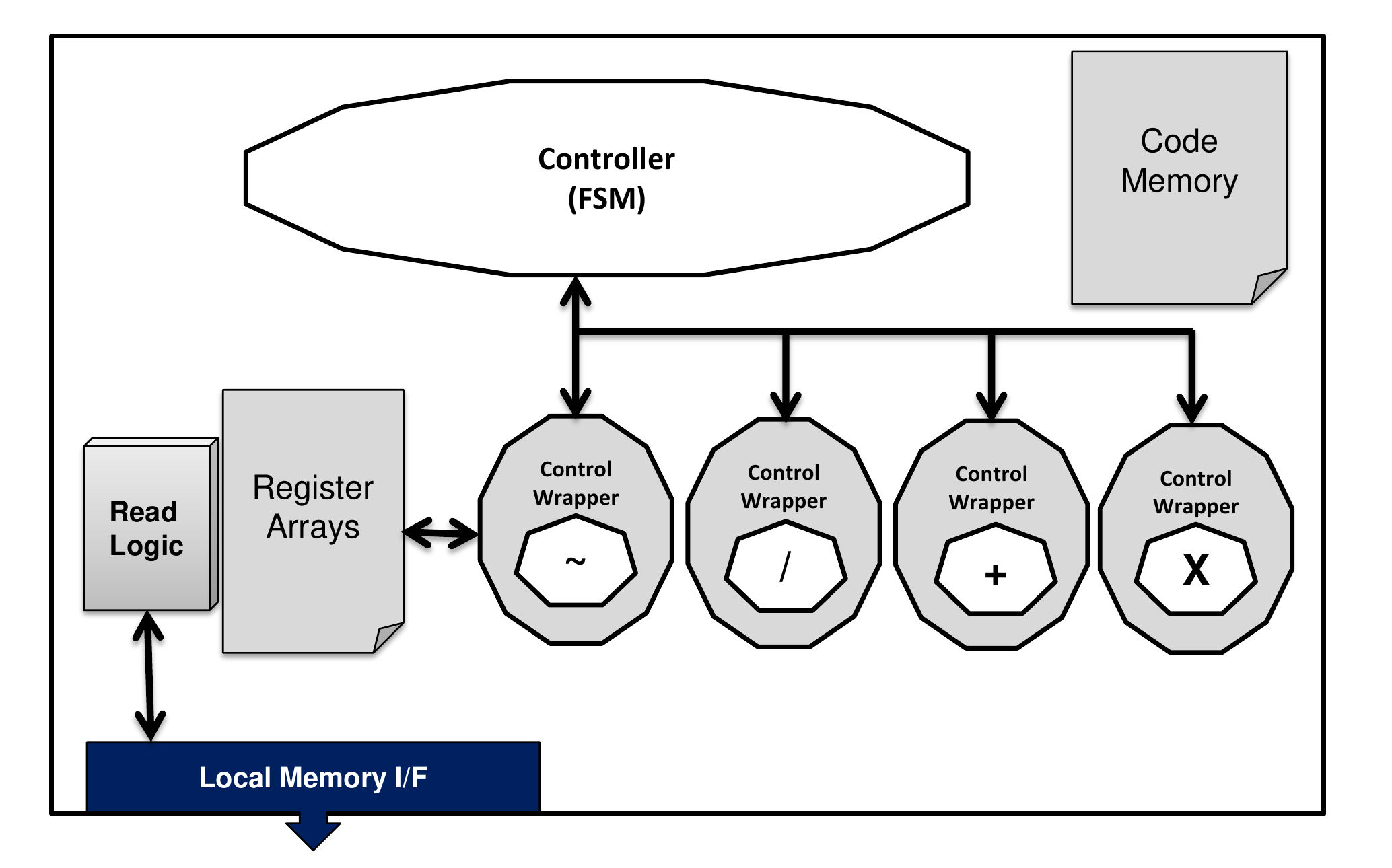}
\caption[]{Second architectural option: fully sequential implementation}
\label{fig:option2}
\end{figure}

One departure from the conventional instruction processor in our core is the additional logic around each arithmetic unit for handling array operations. This allows the main control unit and instructions to work at coarser level of array operations, resulting in a very simple controller and instructions (See Figure \ref{fig:option2}).

%
%
The main advantage of this approach is that it allows very convenient programming of the algorithm. While this particular experiment was for the Convect kernel, the architecture can work as a generic processor. The reconfigurability of the FPGA  allows it to be optimised for each implementation,  similar to approach taken by \cite{112.100}. 

However, this architecture offers no parallelism inside the execution of a single Convect kernel, which is executed in a strictly sequential fashion. There is also the overhead of having a controller around each single arithmetic unit. The result is an architecture that has the overheads of a typical sequential instruction processor, yet would be operating at much lower frequencies than those at which CPUs operate because we are implementing it on an FPGA. 

Our final proposal  builds on the strength of this approach, and  mitigates the overheads by incorporating design elements from the very rich vector-processor design domain.

\subsection{Option 3: Vector Implementation}
\label{sec:vector_implementation}

Based on the first two experiments we propose a more flexible solution. %
Our final architecture introduces vector instructions and multiple arithmetic units.

The convect method, like most scientific applications, requires arithmetic operations on large vectors or arrays. Such applications are very well suited for the SIMD approach, and vector processors have been used successfully in a number of super-computing applications. The Cray computer is a vector-processor \cite{112.027} based on which a number of supercomputing platforms have been built. The Earth Simulator \cite{112.020} at JAMSTECH in Japan is based on a cluster of NEC's NX-9 vector processors.

Our proposed solution implements a Harvard-style vector instruction processor with an array of arithmetic functional units. We introduce vector registers and at the same time we maintain a set of scalar instructions and registers to deal with the non-vector parts of the algorithm, as encountered frequently in a protracted kernel like Convect.

Since our motive has been to evolve towards a generic processing core for scientific computations, we have introduced various levels of flexibility in the design. Compile time parameters can be used to tweak these design aspects: 

\begin{itemize}
\item Optional use of floating-point/fixed-point converter, and fixed-point number representation format.
\item Size of vector registers bank (width and depth).
\item Choice between combinatorial or sequential functional units, and ability to mix both in the same design.
\item Different number of functional units for each arithmetic operation. This is a very useful configuration option as we will show later in our results.
\end{itemize}


We can explore the flexibility along these dimensions with very little effort, and it will be trivial to automate these choices at compile time. This will enable us to make use of a key strength of FPGAs where it can not be matched by either CPUs or GPUs; the compile-time reconfigurability of hardware. Our processor core will only instantiate the hardware required for instructions in a given algorithm known at compile-time. 

\begin{figure}[h]
\centering
\includegraphics[width=1.0\linewidth]{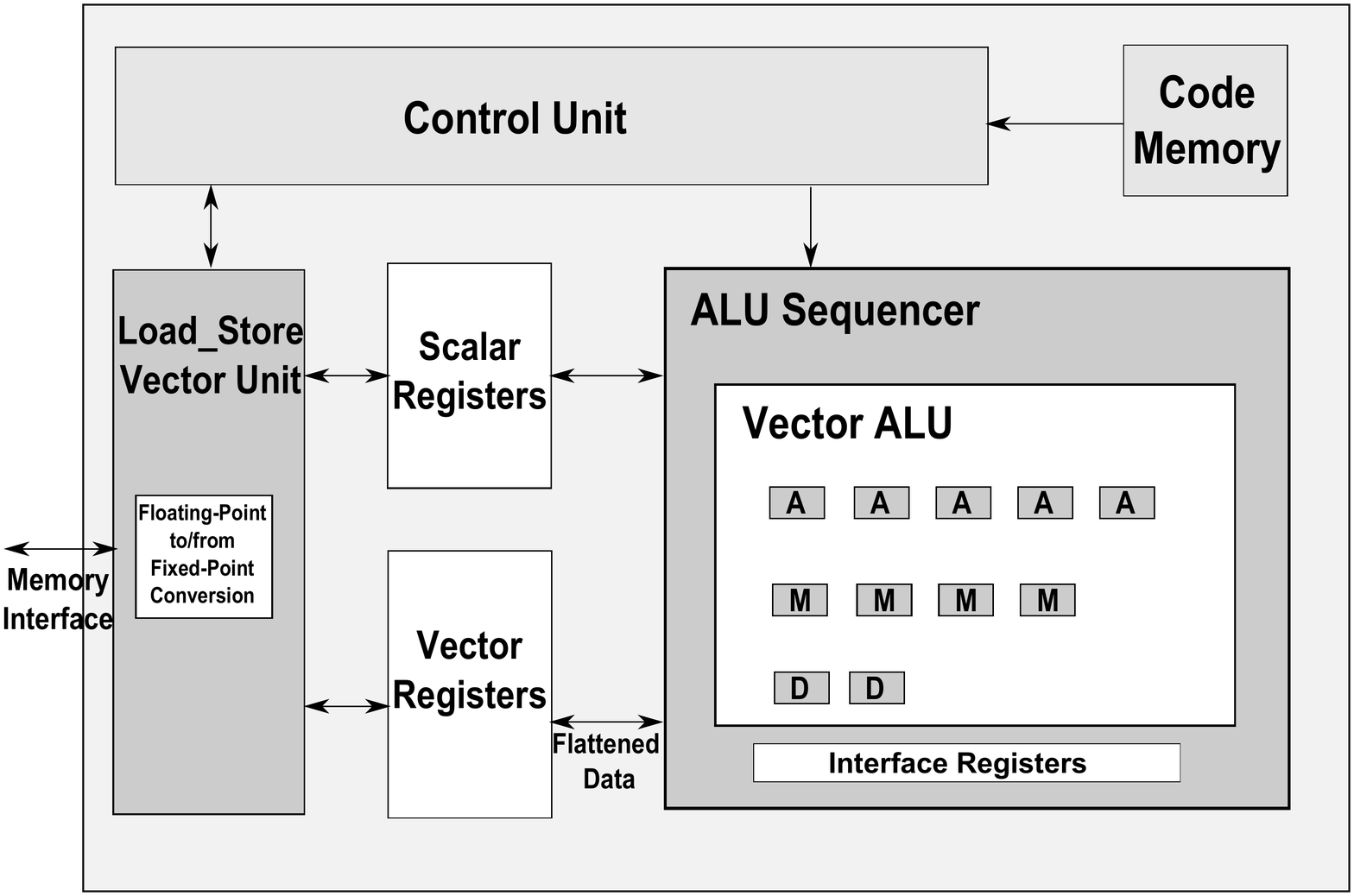}
\caption{Third architecture option: Vector Processing Unit}
\label{fig:option3}
\end{figure}

Figure \ref{fig:option3} shows a block diagram of our proposed architecture. This architecture has been developed in view of a specific problem of executing the Convect kernel, though the instruction set allows generic coding as well. 

The wrapper around the ALU for sequencing the operations is required to deal with the different number of different arithmetic units, which may or may not be the same as the size of the vector. The required sequencing and collection of data is completely transparent at the main control-unit level, and is dealt by the sequencing wrapper which hides this complexity from rest of the architecture as well as from the compiler or programmer.




\section{Resource Utilization and Calculation Latency Comparisons}

In this section we present the results of simulation and synthesis of various architectural options on  Xilinx's XC5VLX devices. We first compare the three architectures that we have presented earlier, and then we explore different configurations of the proposed vector architecture. 

\subsection{Comparing the Three Architectures}
\begin{figure}[h]
\centering
\includegraphics[width=1.0\linewidth]{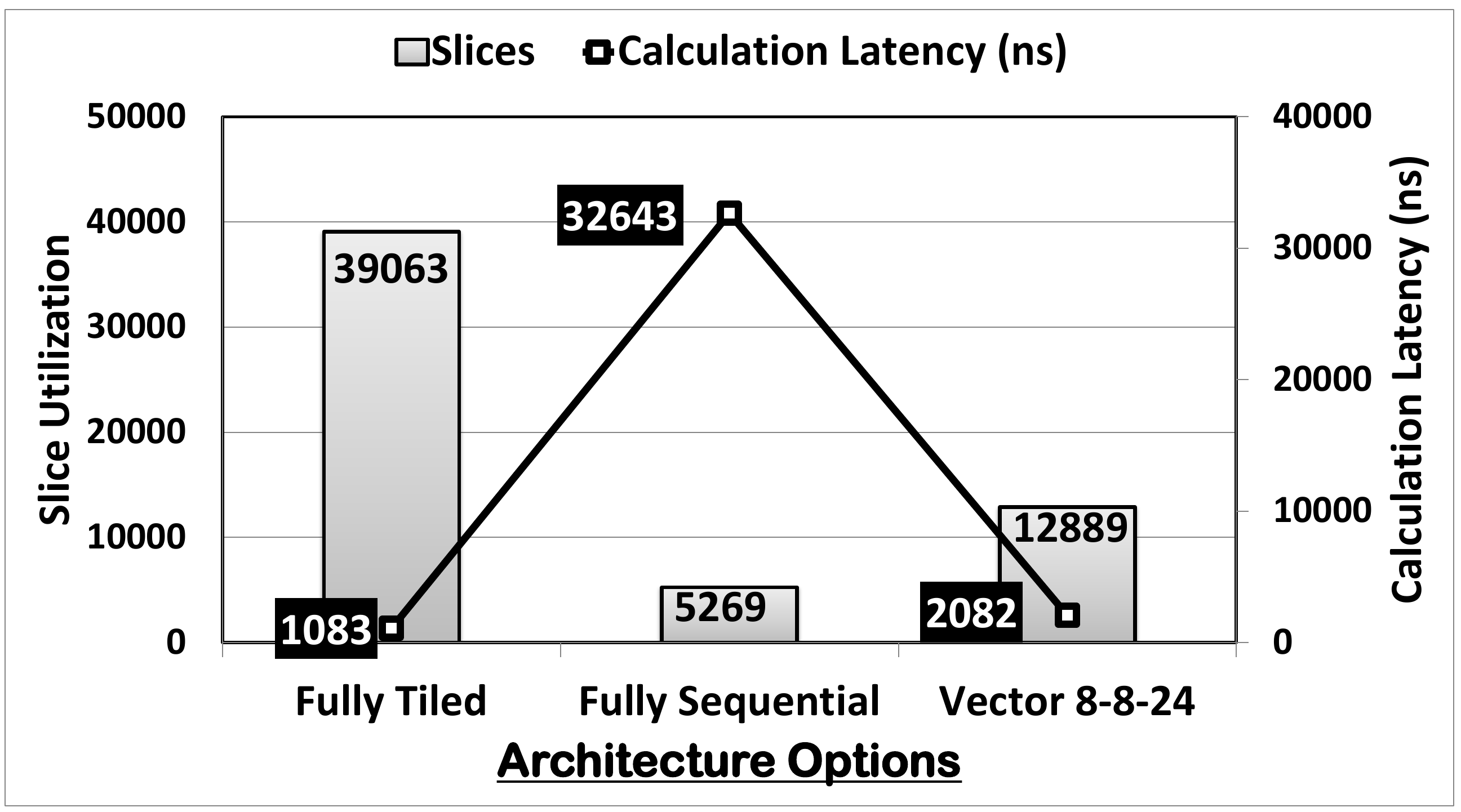}
\caption{Resource utilization and calculation latency comparison of three architectures}
\label{fig:comp_3arch}
\end{figure}

Figure \ref{fig:comp_3arch} shows the comparison of resource utilization and calculation latency of the three architectures presented earlier. For the proposed vector processor solution, we have chosen 8 units each for addition and multiplication, and 24 units for division, which is what 8-8-24 means in the figure. 

The second option of fully sequential implementation consumes about 7$\times$ fewer resources and 8$\times$ more time for the same operation as the fully tiled option 1 (See Figure \ref{fig:comp_3arch}), so there is not much difference between the two in this respect. 

Note however the comparison of the second option (sequential processor)with the third one, that is the vector core, with 8 adders, 8 mulitipliers and 24 dividers. We can see an almost 16$\times$ decrease in latency for a mere 2.5$\times$ increase in the use of slices. Another way of looking at this is that we could instantiate three cores of the first option which make the resource utilization almost the same as the vector processor, yet it will still be almost 6$\times$ slower than the vector processor. This establishes a clear case for the vector processor based architecture.





\subsection{Exploring the Proposed Architecture}

With the flexibility of instantiating different number of arithmetic units in the vector processor core, we experimented with different mixtures of functional units. Note that the divide unit is a sequential one, while adder and multiplier are combinatorial, with the adder having additional logic to allow subtraction as well. The multiplier takes up the most logic resources, while the adder takes up the least resources.

\begin{figure}[h]
\centering
\includegraphics[width=1.0\linewidth]{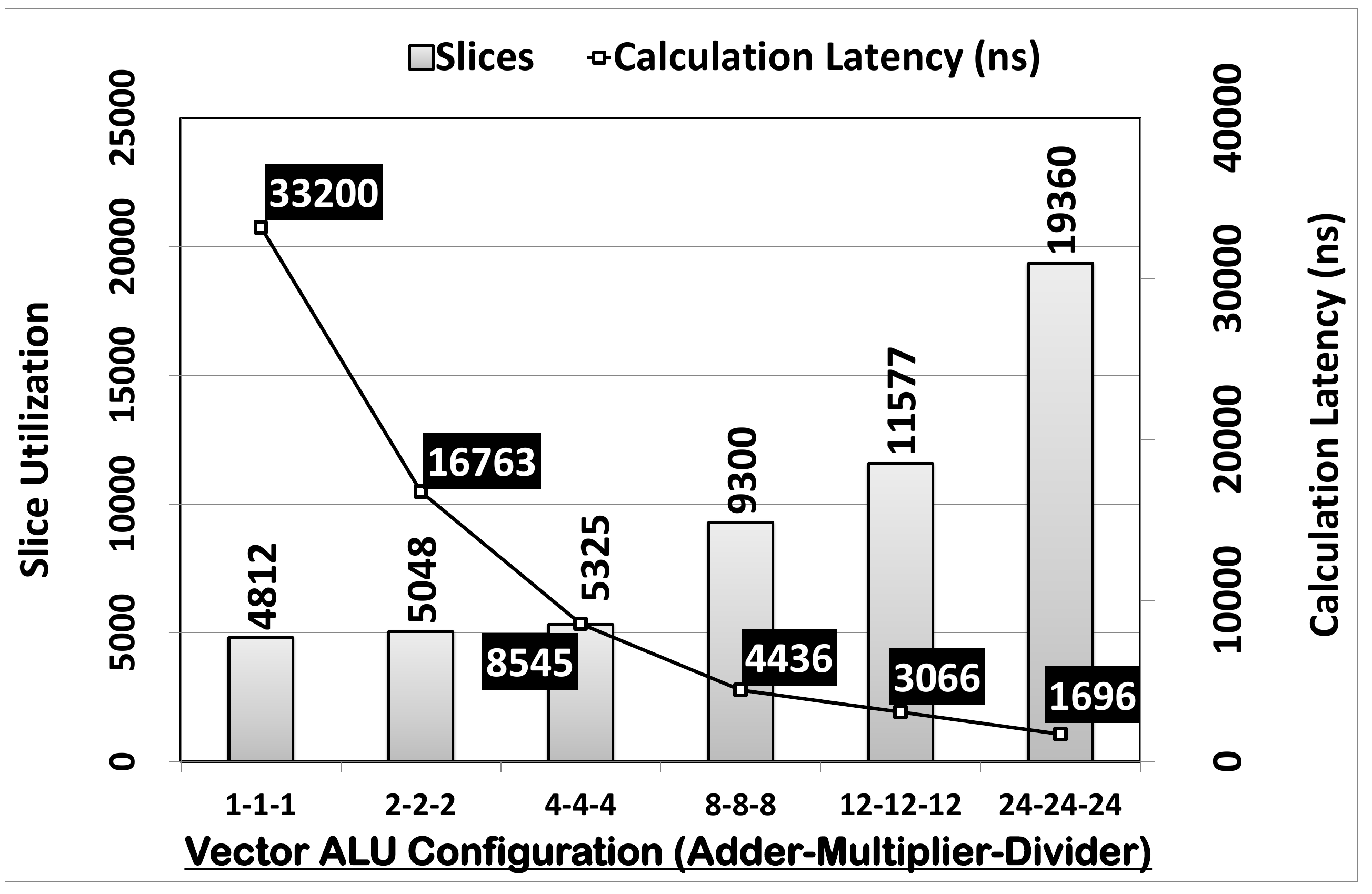}
\caption{Comparing different symmetrical configurations of ALU in the vector architecture}
\label{fig:comp_vector_symmetry}
\end{figure}

Our first set of experiments was on symmetric architectures that keep the number of units for each arithmetic unit equal, so that we can judge the effect of uniformly scaling the vector ALU, from 1 unit each, to one unit each for every vector element. 
The results are outlined in Figure \ref{fig:comp_vector_symmetry}.

The power of the vector architecture can be seen here, where going from a 1-1-1 configuration to a 24-24-24 configuration takes up 4$\times$ more resources, but yields a latency improvement of around 20$\times$. We can also see better yields for resources from a configuration of 4-4-4 and upwards.





\begin{figure}[h]
\centering
\includegraphics[width=1.0\linewidth]{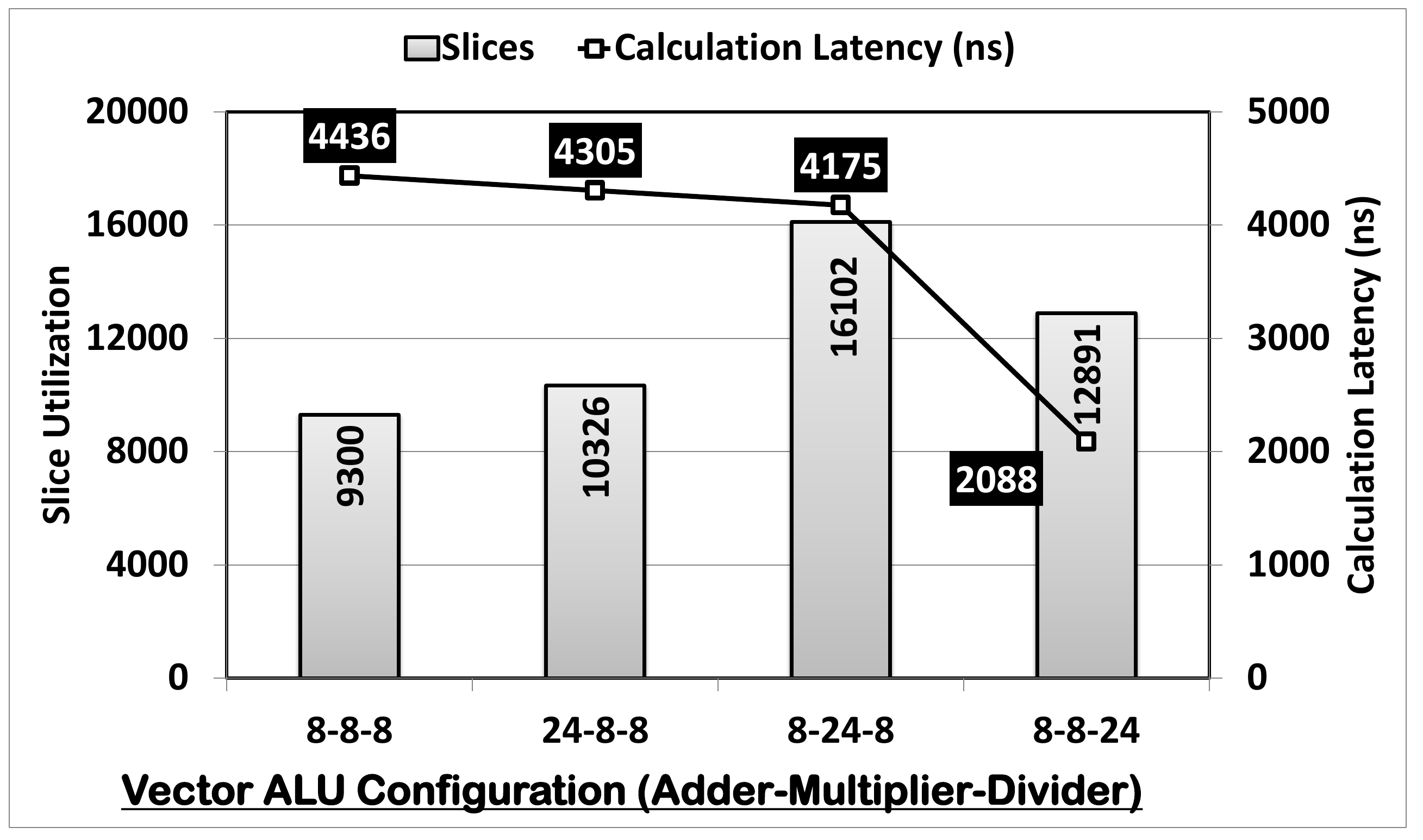}
\caption{Comparing asymmetrical configurations of ALU in the vector architecture against a symmetric 8-8-8 baseline}
\label{fig:comp_vector_8_8_24}
\end{figure}

Next we investigated asymmetrical ALU configurations, by keeping two of the arithmetic unit arrays at 8 instances each, and increasing the third one to see the relative improvements in performance and increase in resources. Our adders and multipliers are pure combinatorial units, whereas the divider is a sequential unit, so one can expect the replication of the divider to yield higher dividends than the other. This expectation is confirmed by our results as shown in Figure \ref{fig:comp_vector_8_8_24}, where the three asymmetrical configurations are compared with each other against the baseline 8-8-8 option. 

Note that increasing adders and especially multipliers increases slice usage considerably, with very little impact on latency, which is largely determined by the sequential unit in the design, that is the divider. 
As expected, the best trade-off is adding dividers which significantly cuts the latency (2.1$\times$)for a relatively modest increase in resource usage (1.4$\times$) when compared with a baseline 8-8-8 configuration. This is the reason we used the 8-8-24 configuration to compare the vector architecture against the other two in Figure \ref{fig:comp_3arch}. 



\section{Discussion, Conclusion and Future Work}

Our observation after experiments with accelerating the Convect kernel in context of accelerating Flexpart has given interesting insights into the subject of acceleration of scientific computations on FPGAs in general. 
We saw that naive approaches like fully unrolling loops and replicating arithmetic units for each and every operation very quickly bloat into an unmanageable size, and would be impractical except perhaps for very simple kernels where the same or a very small set of operations is repeated many times. Using the Convect kernel as a representative algorithm for scientific computations, we can ill afford this approach, especially when one adds the effort required to program a custom circuit for each algorithm.

The other two options both have a major advantage of being programmable in a way very similar to instruction processors. Based on our experiments, our proposal for accelerating scientific models on FPGAs is to use an array of custom vector processors instantiated for the algorithm.

Further experiments with the vector processor indicate the high degree of compile-time configurability of the proposed architecture. Note that all these experiments were done in the context of a particular set of equations, for which we judged the trade-offs. For a different set of equations in another model, some other mix of functional units would yield a different optimal configuration. This aspect of our investigation emphasises the inherent flexibility in the architecture that can allow it to be configured around different kinds of algorithms.

The architecture allows parallelism at three levels:
	\begin{itemize}
	\item At the top (thread) level by replicating the symmetric computational cores, each executing an instance of the function.
	\item It is possible to have a pipeline of asymmetric computational cores, each working on different parts of the algorithm. This has not been explored in case of Convect and is part of our future work.
	\item Data-level parallelism by introduction of multiple arithmetic units in the ALU.
	\end{itemize}

It is a soft processor that can be customised at compile-time to tailor the architecture to the problem.
The computation core has a dedicated code memory whose contents are finalised at compile-time. This means that the core can be customised as per the requirements of the set of instructions it is to execute. 
Programming the architecture at assembly level is quite straightforward as it is essentially a Harvard-style vector processor. While it is not a trivial task, it is certainly possible to take legacy high-level code, partition it to a set of heterogeneous vector cores on an FPGA, and compile the instructions for each core into their code memory, in a fully automated manner. 

In view of the above, we foresee considerable innovation for and contribution to the field of HPC for scientific applications that can be made as we move further in our investigations. 

\bibliography{./literature_universal_WN} 
\bibliographystyle{IEEEtran} 

\end{document}